\documentclass[prd,superscriptaddress,twocolumn,amsmath,amssymb,nofootinbib,longbibliography]{revtex4-1}

\usepackage{graphicx}
\usepackage{amsmath,amssymb}
\usepackage{lmodern}
\usepackage{physics}
\usepackage{color}
\usepackage{slashed}
\usepackage{tensor}
\usepackage[utf8]{inputenc}

\newcommand{\ba}{\begin{eqnarray}}
\newcommand{\ea}{\end{eqnarray}}

\newcommand{\eps}{\varepsilon}

\newcommand{\defterm}[1]{\emph{#1}}              

\newcommand{\tens}[1]{{\ensuremath{\boldsymbol{#1}}}}       

\renewcommand{\grad}{{\tens{d}}}                 
\newcommand{\cv}[1]{{\tens{\partial}}_{#1}}    

\newcommand{\A}[1]{A^{\!(#1)}}                 

\newcommand{\dg}{{{n}}}                           

\newcommand{\KT}[1]{\tens{k}_{(#1)}}                  

\newcommand{\KVc}[1]{l_{(#1)}}                  

\newcommand{\be}{\begin{equation}}             
\newcommand{\ee}{\end{equation}}               
\newcommand{\notprop}{\propto\kern-1\@ptsize pt \diagup}

\begin{document}

\title{Symmetry operators for the conformal wave equation in rotating black hole spacetimes} 


\author{Finnian Gray}
\email{fgray@perimeterinstitute.ca}
\affiliation{Perimeter Institute, 31 Caroline Street North, Waterloo, ON, N2L 2Y5, Canada}
\affiliation{Department of Physics and Astronomy, University of Waterloo,
Waterloo, Ontario, Canada, N2L 3G1}

\author{Tsuyoshi Houri}
\email{t.houri@maizuru-ct.ac.jp}
\affiliation{National Institute of Technology, Maizuru College, Kyoto 625-8511, Japan}

\author{David Kubiz\v n\'ak}
\email{dkubiznak@perimeterinstitute.ca}
\affiliation{Perimeter Institute, 31 Caroline Street North, Waterloo, ON, N2L 2Y5, Canada}
\affiliation{Department of Physics and Astronomy, University of Waterloo,
Waterloo, Ontario, Canada, N2L 3G1}

\author{Yukinori Yasui}
\email{yukinori.yasui@mpg.setsunan.ac.jp}
\affiliation{Institute for Fundamental Sciences, Setsunan University, Osaka 572-8508, Japan}

\date{April 29, 2021}         

\begin{abstract}
We present covariant symmetry operators for the conformal wave equation in the (off-shell) Kerr--NUT--AdS spacetimes. These operators, that are constructed from the principal Killing--Yano tensor, its `symmetry descendants', and the curvature tensor, guarantee separability of the conformal wave equation in these spacetimes. We next discuss how these operators give rise to a full set of conformally invariant mutually commuting operators for the conformally rescaled spacetimes and underlie the $R$-separability of the conformal wave equation therein. Finally,
by employing the WKB approximation we derive the associated Hamilton--Jacobi equation with a scalar curvature potential term and show its separability in the Kerr--NUT--AdS spacetimes.
\end{abstract}

\maketitle

\section{Introduction}
Symmetries, both explicit and hidden, play an important role in general relativity -- in their presence one may be able to explicitly integrate the Einstein equations and/or significantly simplify the study of matter fields in a given curved spacetime.  Perhaps one of the most remarkable symmetries is a hidden symmetry of the principal Killing--Yano tensor ~\cite{Frolov:2017kze}. Such a symmetry appears for the Kerr family of spacetimes in all dimensions, or more precisely for all the so called (off-shell) Kerr--NUT--AdS metrics \cite{Houri:2007xz, Krtous:2008tb, Houri:2008th}, and underlies many of their  remarkable properties. In particular, it stands behind the separability of the massless and massive scalar, spinor, and vector field equations in the Kerr--NUT--AdS backgrounds \cite{Frolov:2006pe, Oota:2007vx, Lunin:2017drx, Frolov:2018ezx}
(see also \cite{Lunin:2019pwz} for a separability of $p$-form fields). 

Most recently, it has been demonstrated \cite{Gray:2020rtr} that also the conformally coupled scalar wave equation
\be\label{CCSF}
\bigl(\Box -\eta R\bigr)\Phi=0\,,\quad \eta=\frac{1}{4} \frac{D-2}{D-1}\,,
\ee
separates in the general off-shell Kerr--NUT--AdS spacetimes. Here, $D$ stands for the number of spacetime dimensions, $R$ is the Ricci scalar of the background metric $\tens{g}$, and prefactor $\eta$ is chosen so that the equation enjoys conformal symmetry, (see, e.g., appendix D of~\cite{wald1984general} ).
Namely, a solution to this equation remains a solution in a conformally scaled spacetime 
\be\label{gt} 
\widetilde{\tens{g}}=\Omega^2\tens{g}\,,
\ee
provided it also scales as $\widetilde{\Phi}=\Omega^{w} \Phi$\,, with the conformal weight $w=1-D/2$\,. The wave equation 
equation \eqref{CCSF} is of fundamental importance and has a number of applications, see e.g. recent study of the asymptotic structure of Kerr spacetime via conformal compactification~\cite{Hennig:2020rns}.

The purpose of the present paper is to further our understanding of the conformal wave equation \eqref{CCSF} in the Kerr--NUT--AdS spacetime ---filling some important gaps in the previous analysis. In particular, we want to `intrinsically characterize' the obtained separability by finding an explicit covariant form of  the corresponding symmetry operators that were found in \cite{Gray:2020rtr} in a given coordinate basis. As we shall see, such operators can be written in terms of the principal Killing--Yano tensor, its symmetry descendants, and the curvature tensor. Moreover, following \cite{Michel:2013dfa}, such operators  can be `lifted up' to conformal operators and guarantee $R$-separability of the conformal wave equation in any conformally related spacetime \eqref{gt}.

Finally, by applying the WKB approximation  we derive an associated with \eqref{CCSF} Hamilton--Jacobi equation with a scalar curvature potential,
\be\label{RHJ0}
 g^{ab}\partial_a S\, \partial_b S+\eta R=0\,,
\ee
and demonstrate its separability in the Kerr--NUT--AdS spacetimes. 
The equation \eqref{RHJ0} has a long history, going back at least to a paper by DeWitt~\cite{DeWitt:1952js} which considers quantum Hamiltonians arising from classical systems. Therein, couplings to the geometrical objects can naturally arise. In a similar vein, the extra term we find in the Hamiltonian can arise due to ambiguities in operator ordering when quantizing non-linear systems~\cite{Omote:1976fx}. It has also found use when considering the quantum mechanics of the motion of a free particle constrained to a Riemannian surface~\cite{Destri:1992sg,Lian:2017qoh}. Here we understand it as a purely classical equation that describes  certain modification of the free particle motion in a curved space.

Our plan for the remainder of the paper is as follows. In the next section we review the Kerr--NUT--AdS spacetimes, their hidden symmetry of the  principal Killing--Yano tensor, and its `symmetry descendants'. In Sec.~\ref{SO} we construct the covariant form of the symmetry operators for the conformal wave equation in these spacetimes. The associated operators for the conformally rescaled metrics are studied in Sec.~\ref{SO2}. In Sec.~\ref{HJ} we derive  the Hamilton--Jacobi equation \eqref{RHJ0} and demonstrate its separability in Kerr--NUT--AdS spacetimes. Sec.~\ref{Con} is devoted to the final discussion. 
Technical results are summarized in Appendices~\ref{AppA} and \ref{AppB}.

\section{Principal Killing--Yano tensor and Kerr--NUT--AdS spacetimes}\label{sec:2}

The {\em principal Killing--Yano} tensor $\tens{h}$ is 
a non-degenerate closed conformal Killing--Yano 2-form $\tens{h}$ obeying the following equation:
\be
\nabla_a h_{bc} = g_{ab}\,\xi_{c} - g_{ac}\,\xi_{b}\,,
\ee
where 
\be\label{xi} 
\xi^a=\frac{1}{D-1}\nabla_bh^{ba}\,
\ee
is the associated primary Killing vector field~\cite{Krtous:2008tb}.

Starting with a single principal Killing--Yano tensor $\tens{h}$, one can generate the whole towers of explicit and hidden symmetries -- the `{\em symmetry descendants}' of $\tens{h}$. In brief, we can construct the following tower of closed conformal Killing--Yano tensors:
\be
\tens{h}^{(j)}=\frac{1}{j!}\underbrace{\tens{h}\wedge \dots \wedge \tens{h}}_{j\ \mbox{\tiny times}}\,.
\ee
Their Hodge duals $\tens{f}^{(j)}=\star\tens{h}^{(j)}$ are Killing--Yano tensors, and their square gives rise to a tower of rank-2 Killing tensors:
\be\label{f2}
k^{ab}_{(j)}=\frac{1}{(D-2j-1)!}f^{(j)a}{}_{c_1\dots d_{d-2j-1}}f^{(j)bc_1\dots c_{d-2j-1}}\,
\ee
for $j\in(0, \dots, n-1)$.  In turn, these tensors give rise to the tower of Killing vectors:
\be\label{Killingcoord}
\boldsymbol{l}_{(j)} = \boldsymbol{k}_{(j)}\cdot \boldsymbol{\xi}^\flat\,.
\ee
Note that the $j=0$ Killing tensor is just the inverse metric and the zeroth Killing vector is the primary Killing vector, $\boldsymbol{l}_{(0)}=\tens{\xi}$. We also have in odd dimensions an extra redundant Killing tensor $\tens{k}_{(n)}=\tens{l}_{(n)}\otimes\tens{l}_{(n)}$.

All of the above constructed symmetries 
mutually Schouten--Nijenhuis commute
\ba
\left[ \boldsymbol{l}_{(i)},\boldsymbol{k}_{(j)} \right]_{\mbox{\tiny SN}}&=&0\;,\; \left[ \boldsymbol{l}_{(i)},\boldsymbol{l}_{(j)} \right]_{\mbox{\tiny SN}}=0\;,\nonumber\\
\left[ \boldsymbol{k}^{(i)},\boldsymbol{k}^{(j)} \right]_{\mbox{\tiny SN}}&=& k^{(i)}_{e(a} \nabla^e k^{(j)}_{bc)} - k^{(j)}_{e(a} \nabla^e k^{(i)}_{bc)} = 0\;\,.
\ea
In addition, the Killing tensors obey the following algebraic identity (i.e. they commute as matrices):
\be\label{KTalg}
k^{a}_{(i)\,b}k^{b}_{(j)\,c}-k^{a}_{(j)\,b}k^{b}_{(i)\,c}=0\,, 
\ee
see \cite{Frolov:2017kze} for all the  details and proofs of the above statements.

The most general spacetime admitting the principal Killing--Yano tensor is the {\em (off-shell) Kerr--NUT--AdS spacetime} \cite{Houri:2007xz, Krtous:2008tb} (see also \cite{Houri:2008th}).
Denoting by ${D=2n+\eps}$ the total number of spacetime dimensions (with $\eps=0$ in even and $\eps=1$ in odd dimensions), the metric takes the following explicit form:
\ba\label{KerrNUTAdSmetric}
\tens{g}&=&\sum_{\mu=1}^n\;\biggl[\; \frac{U_\mu}{X_\mu}\,{\grad x_{\mu}^{2}}
  +\, \frac{X_\mu}{U_\mu}\,\Bigl(\,\sum_{j=0}^{n-1} \A{j}_{\mu}\grad\psi_j \Bigr)^{\!2}
  \;\biggr]\nonumber\\
  && \qquad +\frac{\eps c}{\A{n}}\Bigl(\sum_{k=0}^n \A{k}\grad\psi_k\!\Bigr)^{\!2}\;,
\ea
while the principal Killing--Yano tensor reads
\be
\tens{h}=\sum_{\mu=1}^n \, x_\mu dx^\mu \wedge \left(\sum_{k=0}^{\dg-1}\A{k}_\mu\grad\psi_k\right)\;.
\ee 

The employed coordinates  
$\{x_\mu, \psi_k\}$ have a natural geometrical meaning associated with the principal Killing--Yano tensor. They split into (time and azimuthal angle) \defterm{Killing coordinates} ${\psi_k}$ (${k}={0,\,\dots,\,\dg{-}1{+}\eps}$) that 
correspond to the Killing vectors \eqref{Killingcoord}, 
\be
\boldsymbol{l}_{(k)}=\cv{\psi_k}\,,
\ee
 and the non-trivial (radial and longitudinal angle) coordinates ${x_\mu}$  ($\mu=1,\,\dots,\,\dg$)  that represent the  `eigenvalues' of $\tens{h}$, see \cite{Frolov:2017kze}.

In the above, 
the functions ${\A{k}}$, ${\A{j}_\mu}$, and ${U_\mu}$ are `symmetric polynomials' of the coordinates ${x_\mu}$, and are defined by:
\begin{align}\label{AUdefs}
    \A{k}&=\!\!\!\!\!\sum_{\substack{\nu_1,\dots,\nu_k=1\\\nu_1<\dots<\nu_k}}^\dg\!\!\!\!\!x^2_{\nu_1}\dots x^2_{\nu_k}\;,\:\:\:\nonumber
\A{j}_{\mu}=\!\!\!\!\!\sum_{\substack{\nu_1,\dots,\nu_j=1\\\nu_1<\dots<\nu_j\\\nu_i\ne\mu}}^\dg\!\!\!\!\!x^2_{\nu_1}\dots x^2_{\nu_j}\;,\nonumber\\
U_{\mu}&=\prod_{\substack{\nu=1\\\nu\ne\mu}}^\dg(x_{\nu}^2-x_{\mu}^2)\;,\;\:\: U=\prod_{\substack{\mu,\nu=1\\\mu<\nu}}^n(x^2_\mu-x^2_\nu)=\det\bigl(\A{j}_\mu\bigr)\;,
\end{align}
where we have fixed $\A{0}=1=\A{0}_{\mu}$. Each metric function ${X_\mu}$ is an unspecified function of a single coordinate ${x_\mu}$:
\be
X_\mu=X_\mu(x_\mu)\,.
\ee
Lastly, the constant $c$ only appearing in odd dimensions is a free parameter. 

Despite the fact that the metric is rather complex, its Ricci scalar takes a fairly simple form \cite{Hamamoto:2006zf}
\begin{equation}\label{RSeparated}
    R = \sum_{\mu = 1}^n \frac{r_\mu}{U_\mu}\,,
\end{equation}
where each function $r_\mu$ depends only 
on a single variable $x_\mu$: 
\be\label{rmu}
r_\mu=-X_\mu''-\frac{2\eps X_\mu'}{x_\mu}-\frac{2\eps c}{x_\mu^4}\,.
\ee
The determinant of the metric reads 
\begin{equation}\label{detmetric}
    \sqrt{\abs{g}} = \bigl(c\A{n}\bigr)^\frac{\eps}{2}\, U\,.
\end{equation}

Importantly for our purposes the Killing tensors  $\tens{k}_{(j)}$ take the following coordinate form:
\ba\label{KTjcoor}
  \KT{j}\!
  &=&\!\sum_{\mu=1}^n\! \A{j}_\mu\!\!\left[\! \frac{X_\mu}{U_\mu}\,{\cv{x_{\mu}}^2}\!
  + \!\frac{U_\mu}{X_\mu}\!\left(\!\sum_{k=0}^{n-1+\eps}\!
    {\frac{(-x_{\mu}^2)^{n-1-k}}{U_{\mu}}}\cv{\psi_k}\right)^{\!\!2}\!\right]\nonumber\\
  &&+\eps\,\frac{\A{j}}{c\A{n}}\cv{\psi_n}^2\,,
\ea
where $j=0$ corresponds to the inverse metric, 
$ \tens{g}^{-1}=\KT{0}$.

\section{Separability of the conformal wave equation and its intrinsic characterization}\label{SO}

Recently it was shown \cite{Gray:2020rtr}, that a solution to the conformal wave equation \eqref{CCSF} in the background \eqref{KerrNUTAdSmetric} can be found in the multiplicative separated form,
\begin{equation}\label{ansatz}
    \Phi = \prod_{\mu=1}^n Z_\mu(x_\mu)\prod_{k=0}^{n-1+\eps} e^{i\Psi_k \psi_k}\,,
\end{equation}
where $\Psi_k$ are the Killing vector separation constants, and each of the $Z_\mu$, which is a function of the single corresponding variable $x_\mu$, obeys the following ordinary differential equation:
\ba\label{separated}
{Z_\mu''}&&+{Z_\mu'} \Bigl(\frac{X_\mu'}{X_\mu} +\frac{\eps}{x_\mu}\Bigr) -\frac{Z_\mu}{X_\mu^2}\Bigl(\sum_{k=0}^{n-1+\eps}{(-x_\mu^2)^{n-1-k}}\Psi_k\Bigr)^2\nonumber\\
&&-\frac{Z_\mu}{X_\mu}\Bigl(\eta r_\mu+\frac{\eps}{c x_\mu^2}\Psi_n^2+\sum_{k=0}^{n-1} C_k(-x_\mu^2)^{n-1-k}\Bigr)=0\,,\quad \ \
\ea
where $C_k$ ($k=0,\dots,n-1)$ are the (non-trivial) separation constants and we have set $C_{0}=0$.

As also shown in \cite{Gray:2020rtr}, underlying this separability is a complete set of symmetry operators
$\{{\cal K}_{(j)}, {\cal L}_{(j)}\}$, 
\ba
{\cal K}_{(j)}&=&\nabla_{a} k_{(j)}^{ab} \nabla_{b}-\eta R_{(j)}\,,\label{Qopdef}\\
{\cal L}_{(j)} &=& - i\, \KVc{j}^{a} \nabla_{\!a}\,, \label{Lopdef}
\ea
that all mutually commute one of another,
\be
\label{opcomut}
\bigl[ {\cal K}_{(k)},{\cal L}_{(l)}\bigr]=0\;,\;
\bigl[{\cal L}_{(k)},{\cal L}_{(l)}\bigr]=0\;,\;
\bigl[{\cal K}_{(k)},{\cal K}_{(l)}\bigr]=0\;,
\ee
and one of which is the wave conformal operator. Namely, 
\be \label{K00}
{\cal K}_{(0)}\Psi=0
\ee
is the  conformal wave equation \eqref{CCSF}.
The fact that these commuting operators exist means that, there exists a common eigenfunction of these operators $\Phi$ obeying 
\ba\label{eigenvalue}
{\cal K}_{(j)}\Phi&=&C_j \Phi\,,\\
{\cal L}_{(j)} \Phi&=& \Psi_j \Phi\,.
\ea
It is precisely this eigenfunction which is the separated solution \eqref{ansatz}. 

The operators ${\cal L}_{(j)}$ are the standard scalar operators that are generated from Killing vectors $\boldsymbol{l}_{(j)}$. On the other hand, the Killing tensor operators  ${\cal K}_{(j)}$ pick up, in addition to the standard Killing tensor part $\nabla_{a} k_{(j)}^{ab} \nabla_{b}$, also an `anomalous conformal term' $R_{(j)}$ which ensures the commutation with the conformal wave operator ${\cal K}_{(0)}$. In \cite{Gray:2020rtr} an explicit coordinate expression for this term has been found, it reads
\be\label{Rj}
R_{(j)}=\sum_{\mu=1}^n \frac{A_\mu^{(j)}}{U_\mu} r_\mu\,,
\ee
where $r_\mu$ are the `Ricci scalar functions' \eqref{rmu}. However, no covariant expression for $R_{(j)}$ has been given in \cite{Gray:2020rtr}. 

Here we amend this situation. That is to say, we show in the appendix~\ref{AppA} that $R_{(j)}$ are given in terms of the principal Killing--Yano tensor, its symmetry descendants, and the curvature tensor by the following covariant formula:
\begin{align}\label{eq:CovRj}
R_{(j)}&=k^{ab}_{(j)}R_{ab} +\frac{D-4}{2(D-2)}\Box \Tr(\tens{k}_{(j)})\nonumber\\&\qquad+\alpha_j k_{(j-1)}^{ac}h_c{}^{b}\,(d\xi)_{ab}-\beta_j l_{(j-1)}^a\,\xi_a\nonumber\\
&=k^{ab}_{(j)}R_{ab}+\frac{D-4}{2(D-2)}\Box \Tr(\tens{k}_{(j)})\nonumber\\&\qquad - k_{(j-1)}^{ab}\Bigl(\alpha_jh_a{}^{c}\,(d\xi)_{cb}+\beta_j\xi_a\,\xi_b\Bigr)\;,
\end{align}
where $\tens{\xi}=\tens{l}_{(0)}$ is the primary Killing vector \eqref{xi}, for $j=0$ we defined   
$\tens{k}_{(-1)}\equiv0\equiv\tens{l}_{(-1)}$, and the constants $\alpha_j$ and $\beta_j$ are given by 
\begin{align}
\alpha_j&= \frac{(n-j+\frac{\epsilon}{2})} {(n-1+\frac{\epsilon}{2})}\,\quad
\beta_j=2\frac{(n-j+\frac{\epsilon}{2})}{(n-1+\frac{\epsilon}{2})} (2j - 3)
\,.
\end{align}

Interestingly these objects can be understood as follows. Let us define the following 1-forms $\tens{\kappa}^{(j)}$: 
\begin{equation}\label{eq:kap1}
	\kappa^{(j)}_a= k_{(j)\,a}^{\;\;\;\;\;\;\;b}\nabla_b R\;.
\end{equation}
Then, quantities $R_{(j)}$ can be understood as `potentials' for the above 1-forms: 
\begin{equation}\label{eq:kap2}
	\tens{\kappa}_{(j)}=\tens{d}R_{(j)}\;,
\end{equation}
see appendix \ref{AppA} for the proof. 
In fact, it is this property which underlies the commutation of the operators \eqref{Qopdef}. Given that $[\nabla_{a} k_{(i)}^{ab} \nabla_{b},\nabla_{c} k_{(j)}^{cd} \nabla_{d}]f=0$~\cite{Sergyeyev:2007gf,Gray:2020rtr} for any scalar function $f$, we have
\begin{align}
	\left[\mathcal{K}_{(i)},\mathcal{K}_{(j)}\right]f\!\!\!\!\!\!\!\!&\nonumber\\
	= -\eta&\left(\left[\nabla_{a} k_{(i)}^{ab} \nabla_{b},R_{(j)}\right]f+\left[R_{(i)},\nabla_{a} k_{(j)}^{ab} \nabla_{b}\right]f\right)\nonumber\\
	=-\eta&\left\{\nabla_{a}( k_{(i)}^{ab} \nabla_{b}(R_{(j)}f))-R_{(j)}\nabla_{a}( k_{(i)}^{ab} \nabla_{b}(f))\right.\nonumber\\
	&\left.+R_{(i)}\nabla_{a}( k_{(i)}^{ab} \nabla_{b}(f))-\nabla_{a}( k_{(j)}^{ab} \nabla_{b}(R_{(i)}f))\right\}\nonumber\\
	=-\eta&\left\{
	\nabla_{a}\Big(f\,[k^{ab}_{(i)}\nabla_{b}R_{(j)}-k^{ab}_{(j)}\nabla_{b}R_{(i)}]\Big)\right.\nonumber\\
	+&\left.
	(\nabla_{a}f)\Big(k^{ab}_{(i)}\nabla_{b}R_{(j)}-k^{ab}_{(j)}\nabla_{b}R_{(i)}\Big)\right\}\nonumber\\
	=-\eta&\left\{
	\nabla_{a}\Big(f\,[k^{a}_{(i)\,b}k^{b}_{(j)\,c}-k^{a}_{(j)\,b}k^{b}_{(i)\,c}]\nabla^{c}R\Big)\right.\nonumber\\
	+&\left.
	(\nabla_{a}f)(k^{a}_{(i)\,b}k^{b}_{(j)\,c}-k^{a}_{(j)\,b}k^{b}_{(i)\,c})\nabla^{c}R\right\}\nonumber\\
	=0\ \ &\,,
	\end{align}
where in the final step we have used the algebraic identity \eqref{KTalg}.

We note that, this is a special case of the result presented in~\cite{benenti2002}. Therein, it is shown that the commutation of any operators, $\Box+g$ and $\nabla_{a} K^{ab}\nabla_{b}+f$, where $f,g\in{\cal C}^\infty({\cal M})$ and $K^{ab}$ is a Killing tensor is guaranteed provided
\begin{equation}
\nabla_{a} f=\tensor{K}{_a^b}\nabla_a g-\frac{1}{3}\nabla_{b} (\tensor{K}{_a^c} \tensor{R}{_c^b}-\tensor{R}{_a^c} \tensor{K}{_c^b})\,.
\end{equation}
In the case of the off-shell Kerr--NUT--AdS metrics the final term on the right hand side vanishes as the Killing and Ricci tensors are diagonal in the same basis~\cite{Frolov:2017kze,Hamamoto:2006zf} (See \eqref{eq:RicciT} and \eqref{eq:KTOrtho} in Appendix \ref{AppA}). Thus, this equation reduces to the relationship between \eqref{eq:kap1} and \eqref{eq:kap2}.

\section{Symmetry operators in conformally related spacetimes}\label{SO2}

As mentioned in the introduction, the conformal wave equation \eqref{CCSF} enjoys the conformal symmetry. That is, provided we have a solution $\Phi$ in the spacetime $\tens{g}$, then 
\be\label{phitilde}
\widetilde{\Phi}=\Omega^{w} \Phi\,,\quad w=1-D/2\,
\ee
is a solution of the same equation in the conformally rescaled spacetime 
\be\label{gtilde}
\widetilde{\tens{g}}=\Omega^2\tens{g}\,.
\ee
In particular, this means that \eqref{phitilde} with $\Phi$ given by \eqref{ansatz} yields an {\em $R$-separated solution} of the conformal wave equation in any spacetime related to the off-shell Kerr--NUT--AdS metric by the conformal transformation \eqref{gtilde}.  

It is interesting to ask if also such $R$-separability can be intrinsically characterized by some complete set of mutually commuting operators. In what follows we show that this is indeed the case -- we explicitly construct such operators and discuss their properties. First, starting from the special conformal frame with $\Omega=1$, we scale the operators $\{{\cal K}_{(j)}, {\cal L}_{(j)}\}$, to construct a complete set of mutually commuting operators for the metric $\widetilde{\tens{g}}$, \eqref{gtilde}. Second, 
following \cite{Michel:2013dfa}, we show that such operators can in fact be lifted to conformally invariant operators, providing thus a complete set of conformally invariant mutually commuting operators for the conformal wave equation \eqref{CCSF} in any spacetime related to the Kerr--NUT--AdS metric by a conformal transformation.

\subsection{Mutually commuting operators}
Starting from the mutually commuting operators $\{{\cal K}_{(j)}, {\cal L}_{(j)}\}$ in the special frame with $\Omega=1$, let us define new operators 
$\{\widetilde{\mathcal{O}}_{(j)}, \widetilde{\mathcal{P}}_{(j)}\}$ for general $\Omega$ by:
\ba\label{coobey}
\widetilde{\mathcal{O}}_{(j)}&\equiv&
\Omega^w {\cal K}_{(j)}\Omega^{-w}\,,\nonumber\\ 
\widetilde{\mathcal{P}}_{(j)} &\equiv&
\Omega^w {\cal L}_{(j)}\Omega^{-w}\,.
\ea
By construction such operators mutually commute, as we have 
\begin{align}
\left[\widetilde{\mathcal{O}}_{(i)},\widetilde{\mathcal{O}}_{(j)} \right]&=	\Omega^w\left[\mathcal{K}_{(i)},\mathcal{K}_{(j)} \right]\Omega^{-w}=0\;,\\
\left[\widetilde{\mathcal{O}}_{(i)},\widetilde{\mathcal{{P}}}_{(j)} \right]&=	\Omega^w\left[\mathcal{K}_{(i)},\mathcal{L}_{(j)} \right]\Omega^{-w}=0\,,\\
\left[\widetilde{\mathcal{P}}_{(i)},\widetilde{\mathcal{P}}_{(j)} \right]&=	\Omega^w\left[\mathcal{L}_{(i)},\mathcal{L}_{(j)} \right]\Omega^{-w}=0\,.
\end{align}

Moreover, it follows that when $\Phi$ satisfies the eigenvalue problem \eqref{eigenvalue} in the spacetime $\tens{g}$, $\widetilde{\Phi}=\Omega^w \Phi$ given by \eqref{phitilde} obeys the `associated' eigenvalue problem:
\ba
\widetilde{\mathcal{O}}_{(j)}\widetilde\Phi&=&C_{j}\widetilde{ \Phi}\,,\nonumber\\
\widetilde{\mathcal{P}}_{(j)}\widetilde \Phi&=&\Psi_{j}\widetilde \Phi\,,
\ea
in the conformal spacetime 
$\widetilde{\tens{g}}$. In other words, the operators $\{\widetilde{\mathcal{O}}_{(j)}, \widetilde{\mathcal{P}}_{(j)}\}$, \eqref{coobey}, intrinsically characterize the separability of the conformal wave equation in the conformal spacetime \eqref{gtilde}.

The only `problem' with \eqref{coobey} is that the new operators $\{\widetilde{\mathcal{O}}_{(j)}, \widetilde{\mathcal{P}}_{(j)}\}$ remain expressed in terms of the `old' connection  $\nabla_a$, the old Ricci tensor  $R_{ab}$, and other objects associated with the metric $\tens{g}$ rather than the conformally rescaled metric $\widetilde{\tens{g}}$. However, using the well known transformation properties of the connection and curvature tensor, one can straightforwardly amend this situation. 

For example, let us define the following tilded objects:\footnote{We stress that these objects are not the conformal symmetries of the spacetime $\widetilde{\tens{g}}$, although it is possible to define such symmetries. Namely, the following objects:
\begin{equation*}
k^{ab}_{(j>0)}\,,\quad \Omega^3 h_{ab}\,,\quad
l^a_{(j\geq0)}\,,
\end{equation*}
are the conformal Killing tensors, conformal Killing--Yano 2-form, and conformal Killing vectors of the spacetime $\widetilde{\tens{g}}$. Notice that in doing this, necessarily $\tens{k}_{(0)}=\tens{g}$ transforms differently to the rest of the Killing tensors. One could, of course, use these objects to define the transformed operators, leading to different (seemingly more complex) expressions. We will adopt this strategy for the Killing tensors at least in the next section (\ref{sec: CSO}).
}  
\be
\widetilde k^{ab}_{(j)}=\Omega^{-2} k^{ab}_{(j)}\,,\quad \widetilde h_{ab}=\Omega^2 h_{ab}\,,\quad
\widetilde l^a_{(j)}=\Omega^{-2}l^a_{(j)}\,,
\ee
and raise or lower their indices with the 
metric $\widetilde{\tens{g}}$ and its inverse. We further denote by $\widetilde{\nabla}_a$ the covariant derivative in the spacetime $\widetilde{\tens g}$ and by $\widetilde{R}_{ab}$ its Ricci tensor. With these at hand, the operators \eqref{coobey} can be expressed as follows (see appendix~\ref{AppB} for  details):
\begin{align}
\widetilde{\mathcal{O}}_{(j)}:=&\Omega^2\left(\widetilde{\mathcal{K}}_{(j)}+\eta \left[\Bigl(\widetilde{\nabla}_a\widetilde{\nabla}_b\bigl(\widetilde{k}_{(j)}^{ab}+\frac{1}{2}\widetilde{k}^c_{(j)\,c}\widetilde{g}^{ab}\bigr)\Bigr)\right] \right)\,, \label{eq:ConOps1}\\
\widetilde{\mathcal{P}}_{(j)}:=&\Omega^2\left(\widetilde{\mathcal{L}}_{(j)}-\frac{w}{D-2}\widetilde{\nabla}_a\widetilde{l}^{a}_{(j)}\right)\;,
\label{eq:ConOps2}
\end{align}
where $\widetilde{\mathcal{K}}_{(j)}$ and $\widetilde{\mathcal{L}}_{(j)}$ are given by expressions \eqref{Qopdef}, \eqref{Lopdef}, and \eqref{eq:CovRj}, with all the objects replaced by the tilded ones.
Note that the quantities $\widetilde{\nabla}_b\left[\widetilde{k}_{(j)}^{ab}+\frac{1}{2}\widetilde{k}^c_{(j)\,c}\widetilde{g}^{ab}\right]$ and $\widetilde{\nabla}_a\widetilde{l}^{a}_{(j)}$ vanishes identically when $\Omega=1$ due to the Killing tensor and Killing vector equations
\be
\nabla \tensor[_{(a}]{k}{^{(j)}_b_{c)}}=0\,,\quad \nabla_{\!(a}l^{(j)}_{b)}=0\,,
\ee
respectively.

Moreover, $\widetilde{\mathcal{O}}_{(0)}$ is just a conformally rescaled $\widetilde{\mathcal{K}}_{(0)}$, 
\be\label{K0conf}
\widetilde{\mathcal{K}}_{(0)}=\Omega^{-2} \widetilde{\mathcal{O}}_{(0)}=\Omega^{w-2}\mathcal{K}_{(0)}\Omega^{-w}\,,
\ee
highlighting the conformal invariance of this operator. The other operators, however, take a more complicated form, as is to be expected from the privileged role of the conformal frame with $\Omega=1.$\footnote{This is the only frame where the spacetime admits full (not only conformal) Killing tensors and the Ricci tensor is diagonal in the natural orthonormal frame \cite{Kubiznak:2007kh}.}
We shall return to this issue in the next subsection where we discuss the conformal form of these operators.

\subsection{Conformal symmetry operators}\label{sec: CSO}
Conformal symmetry operators for the conformal wave equation have been studied for many years, see e.g. \cite{carter1977killing,
boyer1976symmetry, kalnins1982intrinsic, kamran1985separation2, duval1999conformally,
benenti2002, eastwood2005higher, eastwood2008higher, gover2012higher, andersson2014second}. This work culminated in 
ref.~\cite{Michel:2013dfa} where a complete and constructive theory was finally formulated. Our goal for the remainder of this section is to review this theory in a more physics community oriented language, and briefly discuss how it applies to the problem at hand.

To start with, we define a conformally invariant operator as an operator that preserves its form under a conformal transformation. More specifically, a conformally invariant operator of weights $s_1$ and $s_2$ obeys the following equality:
\be\label{eq:Opinv}
	\widetilde{Q}_{s_1,s_2}=\Omega^{s_2} Q_{s_1,s_2} \Omega^{-s_1}\,,
\ee
under the conformal transformation \eqref{gtilde}. That is, $\widetilde{Q}_{s_1,s_2}$ has exactly the `same form' as  ${Q}_{s_1,s_2}$ but is constructed out of conformally scaled (tilded) tensors associated with the metric $\widetilde{\tens{g}}$ rather than $\tens{g}$. To give an example,  the conformal wave operator $\mathcal{K}_{(0)}$ obeys the equation \eqref{K0conf} and thence is a conformal operator with weights $s_1=w$ and $s_2=w-2$.

In what follows, we are going to concentrate on conformal operators of equal weights, $s_1=s_2=s$. In particular, as shown in \cite{Michel:2013dfa} the most general second-order  conformal operator with weight $s$ that is built out of a symmetric tensor $K^{ab}$ is given by
\begin{align}\label{eq:GenSymOp}
Q_{s}(K)=&\nabla_{a}K^{ab}\nabla_{b} +\Bigl(\gamma_1[\nabla_{a} K^{ab}]+\gamma_2[\nabla^b \Tr K]\Bigr)\nabla_{b}\nonumber\\
&+\gamma_3(\nabla_{a}\nabla_{b}K^{ab})+\gamma_4(\Box \Tr K)+\gamma_5\,R_{ab}K^{ab}\nonumber\\
&+\gamma_6 \,R\,\Tr K+f\;.
\end{align}
Here $f$ is a function which does not scale under conformal transformation, we assume $\widetilde{K}^{ab}=K^{ab}$, and the  
coefficients are
\begin{align}
\gamma_1&=2\gamma_2=-\frac{(2 s +D)}{D+2}\,,\,\gamma_3=\frac{(s-1) s}{(D+1) (D+2)}\,,\nonumber\\
\gamma_4&=\frac{s (D+2 s-1)}{2 (D+1) (D+2)}\,,\,\gamma_5=\frac{s (D+s)}{(D-2) (D+1)}\,,\nonumber\\
\gamma_6&=-\frac{2 s (D+s)}{(D-2) (D-1) (D+1) (D+2)}\,.
\end{align}
Similarly, having a vector $L^a$, the corresponding conformal operator is given by 
\be\label{eq:Conf KV op}
Q_s(L)=L^{a} \nabla_{\!a} -\frac{s}{D}(\nabla_{\!a}\,L^a)\;.
\ee

In particular, we consider conformal 
operators of weight $w=1-D/2$, c.f. \eqref{phitilde},
\be\label{Qwtilde}
\widetilde Q_w=\Omega^{w} Q_w \Omega^{-w}\,,
\ee that are {\em symmetry} operators of the conformal wave operator $\mathcal{K}_{(0)}$, that is, they satisfy the following relation: 
\begin{equation}\label{SOeq}
\mathcal{K}_{(0)}\circ Q_w={\cal D}\circ\mathcal{K}_{(0)}\,,
\end{equation}
for some operator ${\cal D}$; in fact, it is easy to see that the conformal invariance implies ${\cal D}\equiv{\cal D}_{-2+w}$.  Note that the equation \eqref{SOeq} obviously preserves the kernel of $\mathcal{K}_{(0)}$.

To find such symmetry operators we can use the  following theorem \cite{Michel:2013dfa}: \\
{\bf Theorem 1.} \emph{
Let $K^{ab}$ be a (special) Killing tensor of the metric $\tens{g}$, so that the following conformally invariant `geometric obstruction' built from the Weyl tensor $C_{abcd}$:
\begin{equation}\label{eq:Obs}
\text{Obs}(K)_a=\frac{2(D-2)}{3(D+1)}\Bigl(\nabla_{b}K^{cd}\tensor{C}{^b_c_d_a}-\frac{3}{D-3}K^{cd}\nabla_b\tensor{C}{^b_c_d_a} \Bigr)
\end{equation}
is exact, that is,
\begin{equation}\label{eq:Obsf}
{\bf Obs}(K)=-2 \tens{d}f\,.
\end{equation}
Then \eqref{eq:GenSymOp} with $f$ given by \eqref{eq:Obsf} (up to a constant) is a symmetry operator for the conformal wave operator and in fact satisfies
\begin{equation}
	\mathcal{K}_{(0)}\circ Q_w(K)=Q_{-2+w}(K)\circ\mathcal{K}_{(0)}\,.
\end{equation}
}

When $K^{ab}$ is a Killing tensor we can simplify the operator \eqref{eq:GenSymOp} via the Killing equation,
\be 
\nabla \tensor[_{(a}]{K}{_b_{c)}}=0\,,
\ee
however this will only hold for a particular metric of the conformal class. For this particular metric, we then have 
\ba
Q_w(K)&=&Q_{w-2}(K)\nonumber\\
&=&\nabla_{a} K^{ab}\nabla_{b}
-\frac{(D-2)}{8(D+1)}\left[\Box \Tr K\right]\nonumber\\
&-&\frac{(D+2)}{4(D+1)}R_{ab}K^{ab}+\frac{R\,\Tr K}{2(D+1)(D-1)}+f\;.\quad\quad
\ea
In this case, therefore the corresponding symmetry operator \eqref{SOeq} actually commutes with the conformal wave equation
\be
\left[Q_w,\mathcal{K}_{(0)}\right]=0\,,
\ee
and more generally, we have the conformal commutation
\be\label{dist}
\left[\widetilde{Q}_w,\,\Omega^2\,\widetilde{\mathcal{K}}_{(0)}\right]=0\,,
\ee
valid in any conformal frame.

In particular, taking the Killing tensors 
$\KT{j}$ ($j>0)$ in the Kerr--NUT--AdS metric $\tens{g}$, we find that they satisfy the obstruction condition \eqref{eq:Obsf} with $f_{(j)}$ given by 
\ba\label{eq:ExactObs}
	f_{(j)}&=&\frac{1}{4(1-D^2)}\Bigl[2D\,k^{ab}_{(j)}R_{ab}+3\Box \Tr k_{(j)}\nonumber\\
	&&+(D+1)(D-2)k_{(j-1)}^{ab}\Bigl(\alpha_jh_a{}^{c}\,(d\xi)_{cb}+\beta_j\xi_a\,\xi_b\Bigr)\nonumber\\
	&&-2R\,\Tr k_{(j)}\Bigr]\,.
\ea
It can then easily be checked that\footnote{Of course, the expression \eqref{eq:Obs} is only defined in this coordinate invariant way in the $\Omega=1$ frame although its coordinate expression will be unchanged no matter the frame.} 
the corresponding operators 
\be\label{confK}
{\cal K}_w^{(j)}\equiv Q_w(k_{(j)})\,,
\ee
 \eqref{eq:GenSymOp}, coincide with the operators $\mathcal{K}_{(j)}$, \eqref{Qopdef}, 
\begin{equation}
{\cal K}_w^{(j)}=\mathcal{K}_{(j)}\,.
\end{equation}
Since all of these operators commute with one another for $\Omega=1$, their conformal versions $\widetilde{{\cal K}}_w^{(j)}$, \eqref{Qwtilde} also mutually commute in the spacetime $\widetilde{\tens{g}}$.
Of course, these are nothing else than the operators $\widetilde{\mathcal{O}}_{(j)}$, \eqref{coobey}, this time, however, written 
in a conformally invariant way \eqref{eq:GenSymOp}.\footnote{Although the formulae \eqref{eq:ConOps1} and \eqref{eq:GenSymOp} look rather different, they represent the same operators, and in particular,  the coordinate expressions for the operators $\widetilde{\mathcal{O}}_{(j)}$ and $\widetilde{{\cal K}}_w^{(j)}$ will coincide in any conformal frame. 
The apparent differences arise from how we choose scale the Killing tensors.}
The remaining commutation relations are then guaranteed by 
 \eqref{dist}, since we define for $j=0$
 \be 
\widetilde{{\cal K}}_w^{(0)}\equiv\widetilde{\mathcal{O}}_{(0)}=\Omega^2 \widetilde{\mathcal{K}}_{(0)}\,,
\ee
reflecting the fact that the metric transforms differently than the other Killing tensors under the conformal transformation.

Similarly one can `lift' the operators $\mathcal{L}_{(j)}$, \eqref{Lopdef}, to the conformal ones (as in \eqref{eq:Conf KV op} and c.f. \eqref{eq:ConOps2} where the Killing vectors transform differently)
\begin{equation}
\mathcal{L}^{(j)}_w=-i\, \KVc{j}^{a} \nabla_{\!a} +i\frac{w}{D}(\nabla_{\!a}\,\KVc{j}^a)\;,
\end{equation}
where the second term identically vanishes in the frame $\Omega=1$ where $\KVc{j}^{a}$ are (full, \emph{not} conformal,) Killing vectors. Of course, these will coincide with $\widetilde{\mathcal{P}}_{(j)}$, \eqref{eq:ConOps2}, in any coordinate system.

To summarize, we have found a conformally invariant `generalization' 
$\{\mathcal{K}^{(j)}_w, \mathcal{L}^{(j)}_w\}$ of the symmetry operators \eqref{Qopdef} and \eqref{Lopdef}, with the two being equal in the Kerr-NUT--AdS conformal frame $\tens{g}$. Writing $\widetilde \Phi=\Omega^w \Phi$ in any conformal frame $\widetilde{\tens g}$, these operators obey the following eigenvalue problem:
\ba
\widetilde{\mathcal{K}}^{(j)}_w\widetilde\Phi&=&C_{j}\widetilde \Phi\,,\\
\widetilde{\mathcal{L}}^{(j)}_w\widetilde \Phi&=&\Psi_{j}\widetilde \Phi\,,
\ea
guaranteeing $R$-separability of $\widetilde \Phi$ in any of these frames.

\section{Associated Hamilton--Jacobi equation and its separability}\label{HJ}

We finally turn to study the natural extension of the Hamiltonian--Jacobi equation that arises from the  the geometric optics (WKB) approximation of the conformal wave equation.

Consider the following `$\alpha$-modified conformal wave equation':
\be \label{alpha}
\bigl(\alpha^2\Box -\eta R\bigr)\Phi=0\,.
\ee
Then, upon employing the geometric optics ansatz
\be
\Phi=\Phi_0 \exp\Bigl(\frac{i}{\alpha} S\Bigr)\,,
\ee
while taking the WKB limit $\alpha\to 0$, we arrive at the corresponding Hamilton--Jacobi equation:
\be\label{RHJ}
 g^{ab}\partial_a S \partial_b S+\eta R=0\,.
\ee
This equation is obviously not conformally invariant, however, it is consistent with the particle Hamiltonian,
\be
H=g^{ab}p_ap_b+\eta R\,.
\ee
See e.g.~\cite{DeWitt:1952js,Omote:1976fx} for how such a coupling to the Ricci scalar can arise from quantum corrections.
The equations of motion for this Hamiltonian yield the following modified geodesic equation
\be\label{modGeo}
\frac{D p_a}{d\lambda}=-\eta \partial_a R\,.
\ee

Let us stress that the procedure of deriving \eqref{RHJ} is similar to how one arrives at the massive Hamilton--Jacobi equation starting from the massive ($\alpha$-modified) Klein--Gordon one, e.g. \cite{Sergyeyev:2007gf}. There is, however, a fundamental difference. Namely, the $\alpha$-modified equation \eqref{alpha} is not conformally invariant, unless $\alpha=1$. This is the reason why the WKB limit $\alpha\to 0$ does not produce a conformally invariant Hamilton--Jacobi equation. If instead, one started with the conformal wave equation, setting $\alpha=1$ in \eqref{alpha}, the WKB approximation would then yield the massless Hamilton--Jacobi equation, which of course is conformally invariant.

In what follows we consider the Hamilton--Jacobi equation \eqref{RHJ} of potential physical interest and show its separability in the off-shell Kerr--NUT--AdS spacetimes. Using the form  of the inverse metric given by \eqref{KTjcoor} for $j=0$, the  Hamilton--Jacobi equation \eqref{RHJ} takes the following explicit form: 
\ba
&&\sum_{\mu=1}^n\!\left[\! \frac{X_\mu}{U_\mu}{S_\mu'^2}
+ \frac{1}{U_\mu X_\mu}\!\left(\sum_{k=0}^{n-1+\eps}
(-x_{\mu}^2)^{n-1-k}\!\Psi_k\right)^{\!\!2}\right]\nonumber\\
&&+\eps\,\frac{1}{c\A{n}}\,\Psi_n^2+\eta\sum_{\mu=1}^n \frac{r_\mu}{U_\mu}=0\;,
\ea
where we have used the additive separation ansatz:
\be
S=\sum_{\mu=1}^n S_\mu(x_\mu)+\sum_k \Psi_k \psi_k\,. 
\ee 
Using next the following identity:
\be
\frac{1}{A^{(n)}}=\sum_\mu \frac{1}{x_\mu^2 U_\mu}\,, 
\ee
we can rewrite the previous equation as
\be
\sum_\mu \frac{G_\mu}{U_\mu}=0\,, 
\ee
where 
\be
G_\mu=X_\mu S_\mu'^2+ 
\frac{1}{X_\mu}\!\left(\sum_{k=0}^{n-1+\eps}
(-x_{\mu}^2)^{n-1-k}\,\Psi_k\right)^{\!\!2}+\eps\,\frac{\Psi_n^2}{c x_\mu^2}+\eta r_\mu\,.
\ee
To proceed, we use the separation {\bf Lemma}:
{\em The most general solution of
	\be
	\sum_{\mu=1}^n \frac{f_\mu(x_\mu)}{U_\mu}=0\,,
	\ee
	where $U_\mu$ is defined in \eqref{AUdefs}, is given by
	\be\label{sepfnu}
	f_\mu=\sum_{k=1}^{n-1} C_k(-x_\mu^2)^{n-1-k}\,,
	\ee
	where $C_k$ are arbitrary (separation) constants.}
This yields the following ordinary differential equations for the separated solution:
\ba 
&&X_\mu S_\mu'^2+ 
\frac{1}{X_\mu}\!\left(\sum_{k=0}^{n-1+\eps}
(-x_{\mu}^2)^{n-1-k}\!\Psi_k\right)^{\!\!2}+\eps\,\frac{\Psi_n^2}{c x_\mu^2}+\eta r_\mu \nonumber\\
&&= \sum_{k=1}^{n-1} C_k(-x_\mu^2)^{n-1-k}\,.
\ea

Inverting this expression and identifying the canonical momenta $\tens{p}=\tens{d}S$  the corresponding constants of motion of the modified geodesic equation \eqref{modGeo} are
given by
\be
 C_j=k^{(j)}_{ab} p^a p^b +\eta R_{(j)}\,,
\ee
where $R_{(j)}$ are given by \eqref{eq:CovRj}. It would be interesting to understand what these constants of motion represent physically, e.g. in a quantum system~\cite{DeWitt:1952js,Omote:1976fx}, as this would give a natural interpretation for the functions $R_{(j)}$.

\section{Discussion}\label{Con}

In this paper we have built on the previous work~\cite{Gray:2020rtr} to find covariant forms of the symmetry operators \eqref{Qopdef} and \eqref{Lopdef} of the conformal wave equation in the Kerr--NUT--AdS background \eqref{KerrNUTAdSmetric}. These operators are built out of the  principal Killing--Yano tensor, its symmetry descendants, and the curvature tensor. Moreover their commutativity descends naturally from the commutation properties of the Killing tensors and the special character of the Ricci scalar functions $R_{(j)}$, \eqref{eq:CovRj}. We then showed how to lift these to a full set of conformally invariant mutually commuting symmetry operators $\{{\cal K}_w^{(j)}, {\cal L}_w^{(j)}\}$ that guarantee $R$-separability of the conformal wave equation in any conformally related spacetime $\widetilde{\tens{g}}$, providing thus a highly non-trivial example to the beautiful theory developed in  \cite{Michel:2013dfa}.

The conformal wave equation \eqref{CCSF} is characterized by a specific value of $\eta$. In principle one can consider more general wave equations, where $\eta$ takes any value. It is easy to see that all such equations still separate in the Kerr--NUT--AdS backgrounds; the operators \eqref{Qopdef} and \eqref{Lopdef} commute for any value of $\eta$. However, for general $\eta$ the corresponding wave equations are not conformally invariant and will not separate  in a generic conformally related spacetime. In this case, one could use the conformal properties outlined in appendix \ref{AppB} to construct an equation which separates in the conformal spacetime, however there is no clear physical interpretation for such an equation.

We have also introduced a modified Hamilton--Jacobi equation for a single particle with a Ricci scalar potential term. This equation naturally arises from the WKB limit of the `$\alpha$-modified' conformal wave equation. This limit breaks the conformal invariance and the resulting equation no longer enjoys conformal  symmetry. We have shown that this equation also separates in the Kerr--NUT-AdS spacetimes -- 
the corresponding non-trivial constants of motion are  given by the Killing tensors and the scalar functions $R_{(j)}$, giving a natural setting for the interpretation of the latter.

In future, we would like to study the 
physical implications of the newly derived (non-minimal coupling) Hamilton--Jacobi equation. We also hope to extend the present results to understand separability of conformal fields with higher spin.
 
\section*{Acknowledgements}
\label{sc:acknowledgements}
We would like to thank T. B{\"a}ckdahl for pointing to us the extended  mathematical literature on the conformal wave symmetry operators and in particular the ref.~\cite{Michel:2013dfa}. F.G. acknowledges support from NSERC via a Vanier Canada Graduate Scholarship. Y.Y. is supported by Grant-in-Aid for Scientific Research from the Ministry of Education, Culture, Sports, Science and Technology, Japan No.19K03877.
This work was also supported by the Perimeter Institute for Theoretical Physics and by the Natural Sciences and Engineering Research Council of Canada (NSERC). Research at Perimeter Institute is supported in part by the Government of Canada through the Department of Innovation, Science and Economic Development Canada and by the Province of Ontario through the Ministry of Colleges and Universities. 
Perimeter Institute and the University of Waterloo are situated on the Haldimand Tract, land that was promised to the Haudenosaunee of the Six Nations of the Grand River, and is within the territory of the Neutral, Anishnawbe, and Haudenosaunee peoples.
\\

\appendix

\section{Covariant form of $R_{(j)}$}\label{AppA}
In this appendix we find the covariant form of $R_{(j)}$ in Kerr--NUT--AdS spacetimes by starting from the explicit expressions in canonical coordinates \eqref{KerrNUTAdSmetric}.   
To start with we need an expression for the Ricci tensor. It is rather simple since it is diagonal in the orthonormanl basis of the metric
\begin{align}\label{eq:OrthoB}
\tens{e}^\mu&=\frac{\tens{d}x_\mu}{\sqrt{Q_\mu}}\,,\quad
\hat{\tens{e}}^\mu=\sqrt{Q_\mu}\sum_k A^{(k)}_\mu \tens{d}\psi_k\,,\nonumber\\
\tens{e}^0&=\sqrt{\frac{c}{\A{n}}}\sum_k A^{(k)} \tens{d}\psi_k\,,
\end{align}
where $Q_\mu=X_\mu/U_\mu$. In fact it is given by~\cite{Hamamoto:2006zf,Frolov:2017kze}
\begin{equation}\label{eq:RicciT}
\text{\bf Ric}=-\sum_\mu\hat{r}_\mu\left(\tens{e}^\mu \tens{e}^\mu+\hat{\tens{e}}^\mu\hat{\tens{e}}^\mu\right)-\eps\hat{r}_0\tens{e}^0\tens{e}^0\,,
\end{equation}
where we have introduced
\begin{align}
\hat{r}_\mu&=\frac{\hat{X}_{\mu}''+\frac{\epsilon  \hat{X}_{\mu}'}{x_{\mu}}}{2 U_\mu}+\sum_{\nu\neq\mu}\frac{x_{\nu } \hat{X}_{\nu }'-x_{\mu} \hat{X}_{\mu}'-(1-\epsilon ) (\hat{X}_{\nu }-\hat{X}_{\mu})}{\left(x_{\nu }^2-x_{\mu}^2\right) U_\nu}\;\;,\;\;\nonumber\\ \hat{r}^0&=\sum_\nu\frac{\hat{X}'_\nu}{x_\nu U_\nu}\;,\quad  \hat{X}_\mu=X_\mu+\eps c/x_\mu^2\,.
\end{align}
Also in this basis the Killing tensors are diagonal too,
\be\label{eq:KTOrtho}
\tens{k}_{(j)}
=\!\sum_{\mu=1}^n\! \A{j}_\mu\left[ \tens{e}_\mu \tens{e}_\mu+\hat{\tens{e}}_\mu\hat{\tens{e}}_\mu\right]+\eps\A{j}\tens{e}_0\tens{e}_0\,.
\ee
Hence, using the identity
\begin{equation}
\sum_{\nu\neq\mu}\frac{\A{j}_\mu-\A{j}_\nu}{x^2_\nu-x^2_\mu}=(n-j)\A{j-1}_\mu\;,
\end{equation}
we have
\begin{widetext}
	\begin{align}
	R_{(j)}-k^{ab}_{(j)}R_{ab}&=\sum_\mu\left[\eps\frac{\A{j-1}_\mu x_\mu \hat{X}_{\mu}'}{U_\mu} +2\sum_{\nu\neq\mu}\frac{\A{j}_\mu}{x^2_\nu-x^2_\mu}\left(\frac{x_\mu \hat{X}'_\mu-(1-\eps)\hat{X}_\mu}{U_\mu}+\frac{x_\nu \hat{X}'_\nu-(1-\eps)\hat{X}_\nu}{U_\nu}\right)\right]\nonumber\\
	&=\sum_\mu\left[\eps\frac{\A{j-1}_\mu x_\mu \hat{X}_{\mu}'}{U_\mu}+2\frac{x_\mu \hat{X}'_\mu-(1-\eps)\hat{X}_\mu}{U_\mu} \sum_{\nu\neq\mu}\frac{\A{j}_\mu-\A{j}_\nu}{x^2_\nu-x^2_\mu}\right]\nonumber\\
	&=2\sum_\mu\frac{\A{j-1}_\mu}{U_\mu}\left([n-j+\eps/2]x_\mu \hat{X}'_\mu-(n-j)(1-\eps)\hat{X}_\mu\right)\;.
	\end{align}
	Furthermore, since the Killing vectors satisfy $\nabla \tensor[_{(a}]{l}{^{(j)}_{b)}}=0$ the only information of their derivatives is contained in their exterior derivative. In particular, since in the orthonormal basis \eqref{eq:OrthoB}
	\begin{equation}
	\tens{l}^{(j)}=\sum_{\mu}\A{j}_\mu\sqrt{Q_\mu}\hat{\tens{e}}^\mu+\epsilon\A{j}\sqrt{\frac{c}{\A{n}}}\tens{e}^0\,,
	\end{equation}
	we have that
	\begin{align}
	\tens{d}\tens{l}^{(j)}=&\sum_{\mu}\left[\left(\A{j}_\mu Q_\mu\frac{X_\mu'}{X_\mu}-\eps \frac{2}{x_\mu}\frac{c\A{j}}{\A{n}}+2x_\mu\sum_{\nu\neq\mu}\frac{Q_\mu\A{j}_\mu+Q_\nu \A{j}_\nu}{x^2_\nu-x^2_\mu}\right]\tens{e}^\mu\wedge\hat{\tens{e}}^\mu\right.\nonumber\\
	&\left.+ \eps 2x_\mu\sqrt{Q_\mu}\sqrt{\frac{c }{\A{n}}}\A{j-1}_\mu\tens{e}^\mu\wedge\tens{e}^0+\sum_{\nu\neq\mu}2x_\nu\sqrt{Q_\mu Q_\nu}\frac{\A{j}_\mu-\A{j}_\nu}{x^2_\nu-x^2_\mu}\tens{e}^\nu\wedge\hat{\tens{e}}^\mu\right)\;.
	\end{align}
	Moreover introducing the Killing co-potential
	\begin{equation}
	(D-2j-1)\tens{\omega}^{(j)}_{ab}:=k_{(j)\,a}^{\,\,\,n}h_{nb}=\sum_\mu \A{j}_\mu x_\mu \tens{e}^\mu\wedge \hat{\tens{e}}^\mu\;.
	\end{equation}
	which generates the Killing tensors~\cite{Frolov:2017kze}
	\be
	l^{a}_{(j)}=\nabla_b\,\omega^{ba}_{(j)}\,,
	\ee
	we can calculate
	\begin{equation}
	k_{(j)n}^{\,\,\,a}h^{nb}\,dl_{ab}^{(k)}= 	2\sum_{\mu}\frac{1}{U_\mu}\left(\A{j}_\mu \A{k}_\mu x_\mu \hat{X}'_\mu+\sum_{\nu\neq\mu} \frac{2\hat{X}_\mu(\A{j}_\mu \A{k}_\mu x^2_\mu-\A{j}_\nu \A{k}_\nu x^2_\nu)-\eps c\A{j}_\mu\left(\frac{\A{k}_\nu}{U_\nu}-\frac{\A{k}_\mu}{U_\mu}\right)}{x^2_\nu-x^2_\mu}\right)\;.
	\end{equation}
	Notice the last term proportional to $\eps$ vanishes when $k=0$. Finally let us us calculate $\Box \Tr(\tens{k}_{(j)})$. First, we have
	\begin{equation}
		\Tr(\tens{k}_{(j)})= \eps \A{j}+\sum_\mu 2 \A{j}_\mu=(2(n-j)+\eps)\A{j}\;.
	\end{equation}
	Since this expression only depends on $x_\mu$ we can use the form of the wave operator (see (20) in \cite{Gray:2020rtr}) to write
	\begin{align}
		\nabla_{a}(k_{(j)}^{ab}\nabla_{b} \Tr[\tens{k}_{(j)}])&=\sum_{\mu}\frac{\A{j}_\mu}{U_\mu}\left[X_\mu \partial^2_\mu\Tr(\tens{k}_{(j)})+ \partial_\mu\Tr(\tens{k}_{(j)}) \left( X_\mu'+\frac{\eps}{x_\mu} X_\mu  \right)\right]\nonumber\\
		&=4\sum_{\mu}\frac{\A{j}_\mu\A{j-1}_\mu}{U_\mu}[n-j+\frac{\eps}{2}]\left(x_\mu X'_\mu +(1+\eps)X_\mu\right)\,.
	\end{align}
	Putting this together we have
	\begin{align}
	&\alpha_jk_{(j-1)n}^{\,\,\,a}h^{nb}\,dl_{ab}^{(0)}-\beta_jl_{(j-1)}^a\,l_a^{(0)}+\frac{D-4}{2(D-2)}\Box \Tr(\tens{k}_{(j)})\nonumber\\
	&=2\sum_{\mu}\frac{1}{U_\mu}\left(\A{j-1}_\mu\left(\left[\alpha_j+\frac{(D-4)(n-j+\frac{\eps}{2})}{D-2}\right]x_\mu \hat{X}'_\mu-\left[\frac{\beta_j}{2}-\frac{(D-4)(n-j+\frac{\eps}{2})(1+\eps)}{D-2}\right]\hat{X}_\mu\right)-2\alpha_j\hat{X}_\mu\sum_{\nu\neq\mu} \frac{\A{j}_\mu-\A{j}_\nu}{x^2_\nu-x^2_\mu}\right)\nonumber\\
	&=2\sum_{\mu}\frac{\A{j-1}_\mu}{U_\mu}\left(\left[\alpha_j+\frac{(D-4)(n-j+\frac{\eps}{2})}{D-2}\right]x_\mu \hat{X}'_\mu-\left[\frac{\beta_j}{2}+2(n-j)\alpha_j-\frac{(D-4)(n-j+\frac{\eps}{2})(1+\eps)}{D-2}\right]\hat{X}_\mu\right)\,.
	\end{align}
\end{widetext}
Thus, using $\eps=\{0,1\}$ we can choose the coefficients to be
\begin{align}
\alpha_j&=\frac{2(n-j+\frac{\eps}{2})}{D-2}\;,\\
\beta_j&= \frac{4(n-j+\frac{\eps}{2})}{D-2}(D-3-2(n-j+\frac{\eps}{2}))\,.
\end{align}
Thence we obtain our covariant expression for $R_{(j)}$
\begin{align}
R_{(j)}=&k^{ab}_{(j)}R_{ab}+\frac{D-4}{2(D-2)}\Box \Tr(\tens{k}_{(j)})\nonumber\\
&+\alpha_jk_{(j-1)n}^{\,\,\,a}h^{nb}\,dl_{ab}^{(0)}-\beta_jl_{(j-1)}^a\,l_a^{(0)}\;,
\end{align}
which matches the form in the text upon noting $\tens{l}_{0}=\tens{\xi}$ and $D=2n+\eps$.

Moreover the derivative of $R_{(j)}$ is particularly nice. We can calculate
\begin{align}
\nabla_a R_{(j)}\overset{a=\nu}{=}\sum_{\mu=1}^n \frac{\partial_\nu r_\mu\,A^{(j)}_\mu}{U_\mu}+2x_\nu\A{j}_\nu\sum_{\mu\neq\nu}\frac{\frac{r_\nu}{U_\nu}+\frac{r_\mu}{U_\mu}}{x^2_\mu-x^2_\nu}\nonumber\\
=\frac{r'_\nu\,A^{(j)}_\nu}{U_\nu}+2x_\nu\A{j}_\nu\sum_{\mu\neq\nu}\frac{\frac{r_\nu}{U_\nu}+\frac{r_\mu}{U_\mu}}{x^2_\mu-x^2_\nu}\;.
\end{align}

Notice that one can also construct
\begin{align}\label{eq:sderiv}
k_{(j)\,a}^{\;\;\;\;\;\;\;b}\nabla_b R\,\overset{a=\nu}{=}&\,\sum_{\mu} \A{j}_\nu\,\partial_\nu\frac{r_\nu}{U_\nu}\nonumber\\
&=\frac{ r'_\nu\,A^{(j)}_\nu}{U_\nu}+2x_\nu\A{j}_\nu\sum_{\mu\neq\nu}\frac{\frac{r_\nu}{U_\nu}+\frac{r_\mu}{U_\mu}}{x^2_\mu-x^2_\nu}\nonumber\\
&=\nabla_a R_{(j)}\;.
\end{align}
Thus we have found a covariant expression for our symmetry operators' derivatives
\be\label{eq:gradRj}
\kappa_a^{(j)}:=k_{(j)\,a}^{\;\;\;\;\;\;\;b}\nabla_b R=\nabla_a R_{(j)}\;.
\ee
Clearly $\tens{\kappa}$ is closed and also locally exact in all dimensions thus we can say that our $R_{(j)}$ are the potentials for $\tens{\kappa}^{(j)}$ i.e.
\be
\tens{\kappa}^{(j)}=\tens{d} R_{(j)}\;.
\ee
\newpage

\section{Conformal Transformations}\label{AppB}
Given the spacetime $(\mathcal{M},\tens{g})$ we now consider a conformal transformation of the metric, Killing tensors, and scalar field ($\tens{k}_{(j)}\rightarrow\Omega^{-2}\tens{k}_{(j)}$, $\Phi\rightarrow\Omega^w\Phi$ for $w=1-D/2$) to the conformal spacetime $(\mathcal{M},\tens{g},\Omega)$. The goal of this section is to find a conformally covariant form of our wave operators
\be\label{eq:Waveops}
\bigl(\hat{\mathcal{K}}_{(j)} -\eta R_{(j)}\bigr)\Phi\,,\quad \hat{\mathcal{K}}_{(j)}=\nabla_a k^{ab}_{(j)}\nabla_b\,,\quad \eta=\frac{1}{4} \frac{D-2}{D-1}\,.
\ee

Using the conformal properties of the Ricci tensor and covariant derivatives, we find the following transformations
\begin{align}
\Omega^2\,\hat{\mathcal{K}}_{(j)}\Phi\rightarrow&\nonumber\\
&\Omega^w\left(\hat{\mathcal{K}}_{(j)}+w\nabla_a (k^{ab}_{(j)}\nabla_b\log\Omega)\right.\nonumber\\
&\left. +w(w-2+D)\nabla_a\log\Omega \,k^{ab}_{(j)}\nabla_b\log\Omega \right)\Phi
\end{align}
and
\begin{align}
\Omega^2\,k^{ab}_{(j)}R_{ab}\rightarrow&\nonumber\\
&k^{ab}_{(j)}R_{ab} -\left[(D-2)k^{ab}_{(j)}+k^c_{(j)\,c}g^{ab} \right]\nabla_a\nabla_b\log\Omega\nonumber\\
&+(D-2)\left[k^{ab}_{(j)}-k^c_{(j)\,c}g^{ab} \right] \nabla_a\log\Omega\nabla_b\log\Omega\;.
\end{align}
Thence we have
\begin{widetext}
	\begin{align}
	&\Omega^2\left(\hat{\mathcal{K}}_{(j)}\Phi-\eta k^{ab}_{(j)}R_{ab}\Phi\right)/\Phi\rightarrow\nonumber\\ &(\hat{\mathcal{K}}_{(j)}\Phi-\eta k^{ab}_{(j)}R_{ab}\Phi)/\Phi+w(\nabla_a k^{ab}_{(j)})\nabla_b\log\Omega+((w+\eta(D-2)) k^{ab}_{(j)}{+}\eta k^c_{(j)c}g^{ab})\left[\nabla_a\nabla_b\log\Omega\right]\nonumber\\
	&+(w(w-2+D)-(D-2)\eta) k^{ab}_{(j)}+(D-2)\eta k^c_{(j)c}g^{ab})\left[\nabla_a\log\Omega\nabla_b\log\Omega\right]\nonumber\nonumber\\
	&=\left(\hat{\mathcal{K}}_{(j)}\Phi-\eta k^{ab}_{(j)}R_{ab}\Phi\right)/\Phi+w(\nabla_a k^{ab}_{(j)})\nabla_b\log\Omega-\eta D\,\hat{k}^{ab}_{(j)}\left[(D-2)\nabla_a\log\Omega\nabla_b\log\Omega+\nabla_a\nabla_b\log\Omega\right]\,.
	\end{align}
	Here we have introduced the traceless Killing tensor $\hat{k}^{ab}_{(j)}=k^{ab}_{(j)}-k^c_{(j)\,c} g^{ab}/D$. Clearly this vanishes when $j=0$ so the first operator is conformally invariant. 
	Notice that the last term contains two derivatives of the conformal factor, so consider the term identically zero term (following from the Killing tensor equation) 
	\begin{align}
	\nabla_a\nabla_b\left(k_{(j)}^{ab}+\frac{1}{2}k^c_{(j)\,c}g^{ab}\right)\equiv 0\,.
	\end{align}
	Under the transformation $\tens{k}_{(j)}\rightarrow\Omega^2\tens{k}_{(j)}$ this becomes
	\begin{align}
	\Omega^2\,\nabla_a\nabla_b\left(k_{(j)}^{ab}+\frac{1}{2}k^c_{(j)\,c}g^{ab}\right)&\rightarrow \nabla_a\nabla_b\left(k_{(j)}^{ab}+\frac{1}{2}k^c_{(j)\,c}g^{ab}\right)+ (D+2)(\nabla_a k^{ab}_{(j)})\nabla_b\log\Omega\nonumber\\&+ D\,\hat{k}^{ab}_{(j)}\left[(D-2)\nabla_a\log\Omega\nabla_b\log\Omega+\nabla_a\nabla_b\log\Omega\right]\,.
	\end{align}
	So we have
	\begin{align}
	&\Omega^2\left(\hat{\mathcal{K}}_{(j)}\Phi-\eta\left[ k^{ab}_{(j)}R_{ab}-	\left\{\nabla_a\nabla_b\left(k_{(j)}^{ab}+\frac{1}{2}k^c_{(j)\,c}g^{ab}\right)\right\}\right]\Phi\right)/\Phi\rightarrow\nonumber\\
	&\left(\hat{\mathcal{K}}_{(j)}\Phi-\eta\left[ k^{ab}_{(j)}R_{ab}-	\left\{\nabla_a\nabla_b\left(k_{(j)}^{ab}+\frac{1}{2}k^c_{(j)\,c}g^{ab}\right)\right\} +(D-4)(\nabla_a k^{ab}_{(j)})\nabla_b\log\Omega\right]\Phi\right)/\Phi\,.
	\end{align}
	Note that, as the covariant derivatives and Killing tensors in the second line are in the $\Omega=1$ frame, we have\\$(D-4)(\nabla_a k^{ab}_{(j)})\nabla_b\log\Omega=-(D-4)/2\,(\nabla_a k^{c}_{(j)\,c})\nabla_b\log\Omega$. Thus this term will be canceled by the transformation of $\Box \Tr(\tens{k}_{(j)})$. That is,
		\begin{equation}
		\frac{D-4}{2(D-2)}\Box \Tr(\tens{k}_{(j)})\rightarrow \Omega^{-2}\left(\frac{D-4}{2(D-2)}\Box \Tr(\tens{k}_{(j)}) +\frac{D-4}{2}\nabla_{a} \left[\Tr(\tens{k}_{(j)})\right] \nabla^{a}\log\Omega \right)\,.
		\end{equation}
	
	We now consider the conformal transformation of the final piece;
	\begin{equation}
	{\cal R}_{(j)}:=\alpha_j k_{(j-1)n}^{\,\,\,a}h^{nb}\,dl_{ab}^{(0)}-\beta_j l_{(j-1)}^a\,l_a^{(0)}\,.
	\end{equation}
	Now, if $\tens{k}_{(j)}\rightarrow\Omega^{-2}\tens{k}_{(j)}$ consistency demands that $\tens{h}\rightarrow\Omega^2\tens{h}$ and that $l^a_{(j)}\rightarrow\Omega^{-2}l^a_{(j)}$. That is, one can show on a $p$ form $\star\rightarrow \Omega^{d-2p}\star$. Assuming $h\rightarrow\Omega^rh$; $h^j\rightarrow\Omega^{jr}h^j$ then $f^{(j)}=\star h^j\rightarrow \Omega^{d-4j +jr}f^{(j)}$. So 
	\be
	k^{(j)}_{ab}\propto f_{ac_1\dots c_{D-2j-1}}f_{b}^{c_1\dots c_{D-2j-1}}\rightarrow \Omega^{2(d-4j +jr)+2(D-2j-1)} k^{(j)}_{ab}=\Omega^{2 + 2 j (-2 + r)} k^{(j)}_{ab}\;.
	\ee  
	Hence demanding for all $j$ that $k^{(j)}_{ab}\rightarrow\Omega^2 k^{(j)}_{ab}$ fixes $r=2$. Then, we are left with ${\cal R}_{(j)}$ as a  scalar density of weight $-2$:
	\begin{equation}
	{\cal R}_{(j)}\rightarrow\Omega^{-2}\,{\cal R}_{(j)}\;.
	\end{equation}
	
	Thus putting this all together
	\begin{multline}
	\Omega^2\,\left[\left(\hat{\mathcal{K}}_{(j)} -\eta \left[R_{(j)}-	\left\{\nabla_a\nabla_b\left(k_{(j)}^{ab}+\frac{1}{2}k^c_{(j)\,c}g^{ab}\right)\right\}\right]\right)\Phi\right]/\Phi\rightarrow\\ \left[\left(\hat{\mathcal{K}}_{(j)} -\eta \left[R_{(j)}-\left\{\nabla_a\nabla_b\left(k_{(j)}^{ab}+\frac{1}{2}k^c_{(j)\,c}g^{ab}\right)\right\}\right]\right)\Phi\right]/\Phi\,,
	\end{multline}
	which gives us the form we use in the main text.
\end{widetext}
\bibliographystyle{JHEP}
\bibliography{refs}

\end{document}